\documentclass[aip, jap, amsmath,reprint]{revtex4-1}

\pdfoutput=1  

\usepackage{graphicx}
\usepackage[bookmarksopen=true, colorlinks=true]{hyperref}
\usepackage{color, amssymb}
\usepackage{multirow}

\newcommand{\dd}{\partial}	
\newcommand{\eps}{\varepsilon}	
\newcommand{\s}[1]{_{\rm #1}}

\begin{document}

\title[]{Analytic model of the energy distribution function for highly energetic electrons in magnetron plasmas}

\author{Sara Gallian}
\email{gallian@tet.rub.de}
\author{Jan Tri­e­sch­mann}
\author{Tho­mas Mus­sen­brock}
\author{Ralf Peter Brink­mann}
\affiliation{Institute for Theoretical Electrical Engineering, Ruhr University Bochum,\\ D-44780 Bochum, Germany}
\author{William N. G. Hitchon}
\affiliation{Department of Electrical and Computer Engineering, University of Wisconsin-Madison,\\ Madison, WI 53706, USA}

\date{\today}

\begin{abstract}

This paper analyzes a situation which is common for magnetized technical plasmas such as dc magnetron discharges and HiPIMS systems, where secondary electrons enter the plasma after being accelerated in the cathode fall and encounter a nearly uniform bulk.
An analytic calculation of the distribution function of hot electrons is presented; these are described as an initially monoenergetic beam that slows down by Coulomb collisions with a Maxwellian distribution of bulk (cold) electrons, and by inelastic collisions with neutrals.
Although this analytical solution is based on a steady-state assumption, a comparison of the characteristic time-scales suggests that it may be applicable to a variety of practical time-dependent discharges, and it may be used to introduce kinetic effects into models based on the hypothesis of Maxwellian electrons.
The results are verified for parameters appropriate to HiPIMS discharges, by means of time-dependent and fully-kinetic numerical calculations.
\end{abstract}
\maketitle

\section{Introduction}

Magnetron discharges are mainly sustained by highly energetic electrons that are emitted from the target and accelerated through the sheath acquiring a kinetic energy of about $q_e V\s{b}$. 
A negative bias $V\s{b}$ is applied to the metallic target (or cathode), which emits secondary electrons due to the intense ion bombardment.
These energetic electrons reach the magnetized plasma bulk and cool down both by inelastic collisions 
with neutral species and by interaction with the colder electrons.
All secondary electrons are accelerated by roughly the same potential difference and this acceleration occurs before there is any significant opportunity for inelastic collisions, so the electrons can be treated as a monoenergetic beam. The low pressure of operation of the discharge and the presence of this energetic electron population causes the electron distribution function to be non Maxwellian: a kinetic approach to the electron description is needed.
The dense plasma in the negative glow region, i.e.~the magnetized region close to the target surface, is generated and heated via the interaction with the secondary electron beam. \par
In this work, the majority of the electrons are treated as a Maxwellian population at low temperature (of the order of a few eV). This assumption is supported by results reported in \cite{1991-GuimaraesAlmeidaBretagne} in the study of dc magnetron discharges and by the kinetic global model results reported and discussed in section \ref{AnVsNum}. 
Moreover, experimental measurements performed on dc magnetrons \cite{1991-Sheridan,2004-Seo} show that the eedf is usually well fitted by a Maxwellian in the vicinity of the cathode (the `magnetic trap' region), and by a bi-Maxwellian further away from it. 
In fact, the theory developed in this paper is applicable to the magnetic trap region; here the electrons are trapped in the magnetic field and describe spiral like orbits around the field lines, until they experience a collision.\par
Even though the discharge can be operated at very low pressures (as low as 0.1 Pa), the strong magnetic field effectively confines the electrons and allows them to experience a consistent number of collisions with neutrals, both elastic and inelastic. Therefore a kinetic description of the electrons cannot neglect neutral interactions. \par
The conditions described above are valid both in conventional dc magnetron (dcMS) and in high power impulse magnetron sputtering (HiPIMS) \cite{1999-KouznetsovMacak} systems. 
For a review of HiPIMS discharges see e.g. \cite{2012-GudmundssonBrenning,2011-Anders,2010-SarakinosAlami}. \par
HiPIMS discharges are driven to very high power densities (on the order of kW/cm$^2$) by driving large currents in short pulses: the secondary electrons emitted by the target are accelerated to higher energies with respect to dc magnetron discharges, making the assumption of distinct Maxwellian and energetic electron populations more accurate. Moreover, since the pulse duration is a few hundred microseconds, the electron distribution functions have enough time to reach a steady state configuration. In fact, according to \cite{1964-MontgomeryTidman}, the Maxwellization time is of the order of few tens of nanoseconds (for a plasma density of $5 \cdot 10^{18}~\text{m}^{-3}$ and temperature of 3 eV, it is about 20~ns). The Maxwellization time being one of the fastest time scales, further supports the assumption of describing the cold population with a Maxwellian distribution. The other characteristic time scales of the systems will be discussed in the following paragraphs. 
The key factors to develop a description of the electron distribution function are: the electron acceleration in the cathode fall region, the interaction of the beam electrons with the Maxwellian background by Coulomb collisions and the inelastic collisions with the neutral species.

Recently in \cite{2013-HuoLundin-OhmHeats}, it was claimed that the heating of the electrons takes place in the presheath region as well, and that this contribution is significant in a large variety of cases. 
The model given here neglects heating of the beam by electric fields which might occur after the beam crosses the sheath, but instead assumes the beam acquires an energy of $q_e V\s{b}$ effectively instantaneously. In fact, overall the beam electrons could diffuse far enough across the field lines to gain only a fraction of the presheath voltage, which is about $10$-$20\%$ of $V\s{b}$ \cite{2012-Rauch-PotentialMap}, while slowing down. Therefore any heating mechanism for the energetic electrons taking place outside of the sheath can be neglected.

In this work, a Boltzmann equation for the distribution function of the energetic electrons is given and is solved analytically in a simplified closed form. The influence of the different cooling mechanisms is investigated, and finally the energetic electron distribution is compared with the result of a kinetic global model.


\section{Simplified Boltzmann equation for the energetic electrons}
It is assumed that the electron population is divided into two well separated species: cold or Maxwellian electrons and energetic or hot electrons. 
The Maxwellian population constitutes the majority of the electrons, therefore determining the electron density, and possesses a low temperature in the eV range. 
The energetic (or hot) electrons deposit their energy in the negative glow region and are confined by the magnetic field, until they are slowed down or leave the magnetized region by scattering with both the neutral species and the ions. The plasma is considered to be homogeneous in this region and both electron species are taken to have an isotropic distribution function since the electron/neutral elastic collisions make the distribution spherically symmetric in velocity space.\\
In the paragraphs that follow, the mechanisms responsible for the hot electron cooling are taken into account separately, while their cumulative effect gives the simplified Boltzmann equation for the hot electron energy distribution function (eedf) with [$F^{(h)}$] = m$^{-3}$eV$^{-1}$  
\begin{equation}
	\dfrac{\dd F^{(h)}}{\dd t}  =  \dfrac{\dd J\s{e-e}}{\dd \eps} +  K\s{ion} + K\s{exc} + S - L,
\end{equation}
where the terms on the right hand side describe respectively the electron flux in energy due to Coulomb interactions $\dd J\s{e-e}/\dd \eps$, inelastic ionization $K\s{ion}$ and excitation $K\s{exc}$ collisions. The terms $L$ and $S$ represent the loss and source term of mass and energy to the domain.

\subsection{Coulomb collisions}
In this paragraph the Coulomb interaction of the hot population eedf with the Maxwellian electrons is studied. The hot electron distribution function is allowed to interact only with the cold electron distribution, which is isotropic and Maxwellian with temperature $T\s{eM}$,
\begin{equation}
	f_M(v_M)  =  \dfrac{n\s{e}}{\pi^{3/2} v\s{th}^3} e^{-v_M^2/v\s{th}^2},
\end{equation}
where $n_e$ is the background electron density and $v\s{th}=\sqrt{2 k_b T\s{eM}/m_e}$ is their thermal velocity, with $k_b$ the Boltzmann constant.
In fact, the hot electrons are fast enough and  have a low enough density so as not to interact with themselves, but only the Maxwellian background with $v\s{th} \ll v$. 
Starting from the Fokker Planck equation for isotropic distributions, considering the interaction of $f^{(h)}(v)$ with $f_M$ only, one can write \cite{1964-MontgomeryTidman}
\begin{equation}
	\dfrac{\dd f^{(h)}}{\dd t} = \dfrac{q_e^4 \ln \Lambda}{4 \pi \eps_0^2 m_e^2} n\s{e} \dfrac{1}{v^2} \dfrac{\dd}{\dd v} \left(  f^{(h)} +\dfrac{v\s{th}^2}{2v} \dfrac{\dd f^{(h)}}{\dd v} \right) \label{CC_cgi},
\end{equation}
where $q_e$ and $m_e$ are the electron charge and mass, $\ln \Lambda$ is the Coulomb logarithm and $f^{(h)}$ is normalized so that the hot electron density is $n^{(h)} = \int f^{(h)}(v) 4 \pi v^2 dv$. 
Here the plasma parameter is defined as $\Lambda = b_\perp/\lambda\s{D}$, where $\lambda\s{D}$ is the Debye length  due to the Maxwellian electrons, and $b_\perp$ is the impact parameter $b_{\perp} = q\s{e}^2/(2\pi\epsilon_0 m\s{e} v^2\s{th})$.
The first term in brackets in equation \eqref{CC_cgi} represents a drift in velocity space, while the second is a diffusion. 
This expression can be rewritten in terms of the slowing down frequency $\nu_s^{hM}$ and parallel velocity diffusion $\nu_\parallel^{hM}$ \cite{2005-Helander} as
\begin{equation}
	\dfrac{\dd f^{(h)}}{\dd t} = \dfrac{1}{v^2} \dfrac{\dd}{\dd v} \left[  \dfrac{v^3}{2} \left(\nu_s^{hM} f^{(h)} + \nu_\parallel^{hM} v \dfrac{\dd f^{(h)}}{\dd v} \right)  \right], \label{CC_SI}
\end{equation}
where the superscript $'hM'$ refers to the interaction of electrons of species $h$, i.e. hot electrons, with Maxwellian electrons $M$.
The frequency $\nu_s^{hM}$ represents the rate at which the hot electrons are decelerated by collision with the cold ones and in the case of energetic electrons interacting with Maxwellian ones can be written as
\begin{equation}
	\nu_s^{hM}(\eps) = \dfrac{q_e^4 \ln \Lambda}{4 \pi \eps_0^2 m_e^2} n\s{e} \dfrac{2}{v^3}. \label{nushM}
\end{equation}
Since $v\s{th} \ll v$, one can neglect the diffusion term, or the parallel velocity diffusion term. It is convenient to express equation \eqref{CC_SI} with $f^{(h)}(v)$ in terms of the kinetic energy $\eps = \frac{1}{2} m_e v^2/q_e$  in eV, and to normalize the distribution function accordingly as $n^{(h)} = \int F^{(h)}(\eps) d\eps$. 
Therefore the eedf $F^{(h)}$ is related to $f^{(h)}$ by \cite{2005-LiebermanLichtenberg-pg191}
\begin{equation}
	F^{(h)}(\eps) = 2\pi \left(\dfrac{2 q_e}{m_e}\right)^{3/2} \sqrt{\eps} \ f^{(h)}(v).
\end{equation}
Then the evolution of $F^{(h)}(\eps)$ due to the interaction with a Maxwellian electron background is given by
\begin{equation}
	\dfrac{\dd F^{(h)}}{\dd t}  =  \dfrac{\dd}{\dd \eps} \left( \eps \nu_s^{hM}  F^{(h)} \right).  \label{CC_eps}
\end{equation}
It will be shown below that the solution has a very weak dependence on $v$, further helping to justify the neglect of the diffusive term.

\subsection{Inelastic collisions}
In this paragraph inelastic collision processes are taken into account. First the ionization term is investigated, then that for excitation.  
Following \cite{1981-BretagneDelouya}, the ionizing collision term for an isotropic distribution function has the integral form
\begin{eqnarray}
	K\s{ion}(\eps, t) =&& \, n_n(t) \int_{\eps\s{ion}+\eps}^{\infty} v(\eps ') \ \sigma\s{ion}(\eps ',\eps) f(\eps ',t) \ d\eps' \nonumber \\
	- &&\, n_n(t) H(\eps-\eps\s{ion}) \sigma_{\rm ion}(\eps) v(\eps) f(\eps,t), \label{Kion}
\end{eqnarray}
where the positive term gives the contribution of incident electrons with energy $\eps'$ larger than $\eps\s{ion}+\eps$ that had an ionizing collision and produced an electron with energy $\eps$. The probability of an ionizing collision that produces an electron at $\eps$ when the incident energy was $\eps'$ is given by the differential cross section $\sigma\s{ion}(\eps ',\eps)$. The negative term refers to the incident electron having energy $\eps$. Here $n_n(t)$ is the gas density, $H(x)$ the Heaviside step function and $\eps\s{ion}$ the ionization threshold.
By definition, the dimension of the differential cross section is $[\sigma\s{ion}(\eps ',\eps)] =$ m$^2$/eV whereas the dimension of the total cross section is $[\sigma\s{ion}(\eps)] =$ m$^2$. The two quantities are related by
\begin{equation}
	\sigma\s{ion}(\eps) = \int_0^{(\eps-\eps\s{ion})/2} \sigma\s{ion}(\eps,\eps_s) \ d\eps_s, \label{TotCS}
\end{equation}
where $\eps_s$ is the scattered electron energy, and the energy $\eps-\eps\s{ion}$ is the residual energy of the incident electron after an ionizing collision. 
Alternatively $\sigma\s{ion}(\eps,\eps_s)$ can be defined by introducing a function $g_{h}(\eps', \eps)$
\begin{equation}
	\sigma\s{ion}(\eps') g_{h}(\eps', \eps) = \sigma\s{ion}(\eps', \eps), \ \int_0^{\eps'-\eps\s{ion}} g_{h}(\eps', \eps) d\eps = 2.
\end{equation}
Given the considerable separation on the energy scale of cold and hot electrons, it is assumed that a hot electron with incident energy $\eps'\gg\eps\s{ion}$ experiencing an ionizing collision will produce a cold electron at zero energy and a hot electron at the residual energy $\eps\s{r} = \eps'-\eps\s{ion}$. Expressing this mathematically
\begin{equation}
	g_{h}(\eps', \eps) = \delta(\eps) + \delta(\eps-\eps\s{r}).
\end{equation}
Substituting into equation \eqref{Kion}, gives
\begin{eqnarray}
	\left. K\s{ion} \right|_h = && \, n_n(t) \int_{\eps\s{ion}+\eps}^{\infty} F^{(h)}(\eps',t) \sigma_{\rm ion}(\eps') \delta(\eps-\eps\s{r}) v(\eps ')   \ d\eps' \nonumber \\
	- && \, n_n(t) F^{(h)}(\eps,t) \sigma_{\rm ion}(\eps) v(\eps) \nonumber \\
	= && \, F^{(h)}(\eps+\eps\s{ion},t) \nu\s{ion}(\eps+\eps\s{ion},t) \nonumber \\
	- &&\,  F^{(h)}(\eps,t) \nu\s{ion}(\eps,t).
\end{eqnarray}
where the ionization frequency $\nu\s{ion}(\eps,t) = n_n(t) \sigma_{\rm ion}(\eps) v(\eps)$ was introduced. Since hot electrons possess large energies $\eps \gg \eps\s{ion}$, one can approximate
\begin{equation}
	\left. K\s{ion} \right|_h \approx \eps\s{ion} \dfrac{\dd  }{\dd \eps}( \nu\s{ion} F^{(h)} ).
\end{equation}
Repeating the same steps, the excitation term can be written as
\begin{equation}
	\left. K\s{exc} \right|_h \approx \sum_p \eps^{(p)}\s{exc} \dfrac{\dd}{\dd \eps} \left( \nu^{(p)}\s{exc} F^{(h)} \right),
\end{equation}
under the assumption $\eps \gg \eps^{(p)}\s{exc}$, where the superscript $(p)$ represents the $p$-th excitation level.

\subsection{Loss term and boundary condition}\label{Bcs}
So far, the derivation of the Coulomb and inelastic collision terms is rather general and can be applied to all systems showing an energetic low density electron population interacting with a high density Maxwellian one. The loss term written in terms of scattering of the electrons out of the volume under consideration is peculiar to planar magnetron systems. Moreover, the boundary condition which introduces hot electrons with energy $q_e V_b$ is specific to magnetron systems (both in dc and HiPIMS mode).
The HiPIMS case is addressed for the following reasons: the existence of a low temperature dense Maxwellian electron population is ensured by the high density of the plasma; the secondary electrons emitted by the target are highly energetic since the bias voltage (and therefore the sheath potential) can be larger than the typical dc case.
The hot electron loss term in a magnetron is written hereafter \cite{1991-GuimaraesAlmeidaBretagne} as
\begin{equation}
	L(\eps) = \frac{1}{4} \frac{A_{\rm L}\ r_{\rm L}}{V}  \left( \nu^{\rm tot}\s{e-n} + \nu\s{ei} \right) F^{(h)}, \label{Loss}
\end{equation}
where $A_{\rm L}$ is the loss area, $r\s{L}$ is the Larmor radius, $\nu^{\rm tot}\s{e-n}$ is the total collision frequency of the electrons with the neutral species, and $\nu_{\rm ei}$ is the Lorentz scattering collision frequency for electron-ion Coulomb collisions \cite{2005-Helander}:
\begin{equation}
	 \nu\s{ei}(\eps) = \frac{q_e^4 \ln \Lambda}{4 \pi m_{\rm e}^2 \eps_0^2} n\s{i} \frac{1}{v^3}.
\end{equation}
The introduction of electrons with the correct mass flow and energy is described by a term in the kinetic equation given by \cite{1991-GuimaraesAlmeidaBretagne}
\begin{equation}
	S(\eps) = \gamma\s{sec}\frac{I\s{D}}{V\ q_e} \delta(\eps-q_e V_b).
\end{equation}
Here $\gamma\s{sec}$ is the secondary electron emission coefficient, $I\s{D}$ the discharge current, $V$ the region volume, and $V_b$ the bias applied at the target.
This term has strictly the form of a boundary condition on the energy axis at $\eps =  q_e V_b$ and it represents the number density of electrons per unit time and energy that enters the volume after the acceleration in the sheath up to an energy of $q_e V_b$.


\section{Kinetic equation for the hot electrons}\label{Fh_sol}
Bringing the Coulomb and the inelastic collision terms together, the equation for the evolution of the hot electrons' eedf reads
\begin{equation}
\begin{split}
	\dfrac{\dd F^{(h)}}{\dd t} =& \,  \dfrac{\dd}{\dd \eps} \left( \eps \nu_s^{hM}  F^{(h)} \right) + \sum_p \dfrac{\dd}{\dd \eps} \left( \eps^{(p)}\s{exc}  \nu^{(p)}\s{exc} F^{(h)} \right)\\
	 + &\, \dfrac{\dd}{\dd \eps} \left( \eps\s{ion} \nu\s{ion} F^{(h)} \right) 	+ \gamma\s{sec}\frac{I\s{D}}{V q_e} \ \delta(\eps-q_e V_b) \\
	 - &\, \frac{1}{4} \frac{A_{\rm L}\ r_{\rm L}}{V}  \left( \nu^{\rm tot}\s{e-n} + \nu\s{ei} \right) F^{(h)}. \label{Fhsimpl2}
\end{split}
\end{equation}
In steady state, equation \eqref{Fhsimpl2} can be written as
\begin{equation}
	\dfrac{d}{d \eps} \left(v_\eps(\eps) F^{(h)}\right) - \nu^*\s{L}(\eps) F^{(h)} + S \delta(\eps -q_e V_b) = 0, \label{Fhss}
\end{equation}
where $S=\gamma\s{sec}{I\s{D}}/{(V q_e)}$ is a constant, $v_\eps(\eps)$ represents a drift velocity in energy space (a measure of the cooling speed of the hot electrons), and $\nu^*\s{L}$ a geometry weighed equivalent loss frequency. $v_\eps(\eps)$, strictly the energy loss velocity, is defined as
\begin{equation}
v_\eps(\eps) =\eps^{(p)}\s{exc} \sum_p \nu^{(p)}\s{exc}  +\eps\s{ion} \nu\s{ion}  + \eps \nu\s{s}^{hM}, \label{veps}
\end{equation}
and $\nu^*\s{L}(\eps)$ is  
\begin{equation}
	\nu^*\s{L}(\eps) = \frac{1}{4} \frac{A_{\rm L}\ r_{\rm L}}{V}  \left( \nu^{\rm tot}\s{e-n} + \nu\s{ei} \right). 
\end{equation}
The sensitivity of $v_\eps(\eps)$ to the electron and neutral densities is addressed in section \ref{SensitivityAnalysis}. 
Equation \eqref{Fhss} can be solved by using the integrating factor technique, and gives
\begin{equation}
	F^{(h)}(\eps) =\dfrac{S}{v_\eps(q_e V_b)} \dfrac{M(q_e V_b)}{M(\eps)} , \label{Fh}
\end{equation}
where $M$ is the integrating factor,
\begin{equation}
	M(\eps) = \exp \left( \int_{q_e V_b}^\eps \dfrac{v'_\eps - \nu^*\s{L}}{v_\eps} d\eps' \right), \label{Meps}
\end{equation}
with $v_\eps' = dv_\eps/d\eps$. 
The assumption of steady state signifies that the hot electrons do not have direct memory of the discharge evolution, but they respond instantaneously to the input quantities, e.g. the species densities.

\subsection{Approximate kinetic equations for perfectly confined electrons}
It is interesting to simplify further equation \eqref{Fhsimpl2}, under the assumption that the electrons are perfectly confined by the magnetic field, and leave the magnetized region only after having deposited their energy in the negative glow region. The steady state equation \eqref{Fhss} reduces to
\begin{equation}
	\dfrac{d}{d \eps} \left(v_\eps(\eps) F^{(h)}\right) = S \delta(\eps -q_e V_b), \label{Fhss_veps}
\end{equation}
with solution 
\begin{equation}
	F_{LL}^{(h)} =  \dfrac{S}{v_\eps(\eps)}. \label{Fh_noLoss}
\end{equation}
This approximation may be considered as appropriate if the electron losses are negligible, i.e. the magnetic field strength is very large.


\section{Analytic solutions versus numerical results}\label{AnVsNum}

The expressions for $F_{LL}^{(h)}$ and $F^{(h)}$ are specified for the HiPIMS system described in \cite{2012-EhiasarianHecimovic} by relying on calculated physical parameters (i.e. the species densities).
These values are calculated by employing a zero dimensional global (or volume averaged) model, which will be described in detail in a follow up publication. This global model evolves the eedf together with the densities of other species, namely: argon and aluminum neutral atoms, argon metastable species, single charge argon and aluminum ions and double charge aluminum ions. The electrons are heated in the same fashion described in { section} \ref{Bcs} by the insertion of secondary electrons at the energy corresponding to the bias voltage applied to the cathode. This bias voltage is given as input and it is kept constant during the on time of the pulse (400 $\mu$s). The discharge current is calculated self consistently from the ion fluxes to the cathode, as well as the metal (Al) production by sputtering and self sputtering. 
The discharge conditions are: bias voltage 360 V, Ar pressure 0.5 Pa, average magnetic field in the region 80 mT and magnetron diameter 50 mm. Following \cite{2011-RaaduAxnaes}, the secondary electron emission coefficient is set to about 0.08 \cite{1983-Yamauchi}, while the self sputtering and sputtering coefficients are estimated from \cite{1969-Hayward,2007-Anders-VICharact,1990-Ruzic}.

\begin{figure}[!h]
\includegraphics[width=8cm]{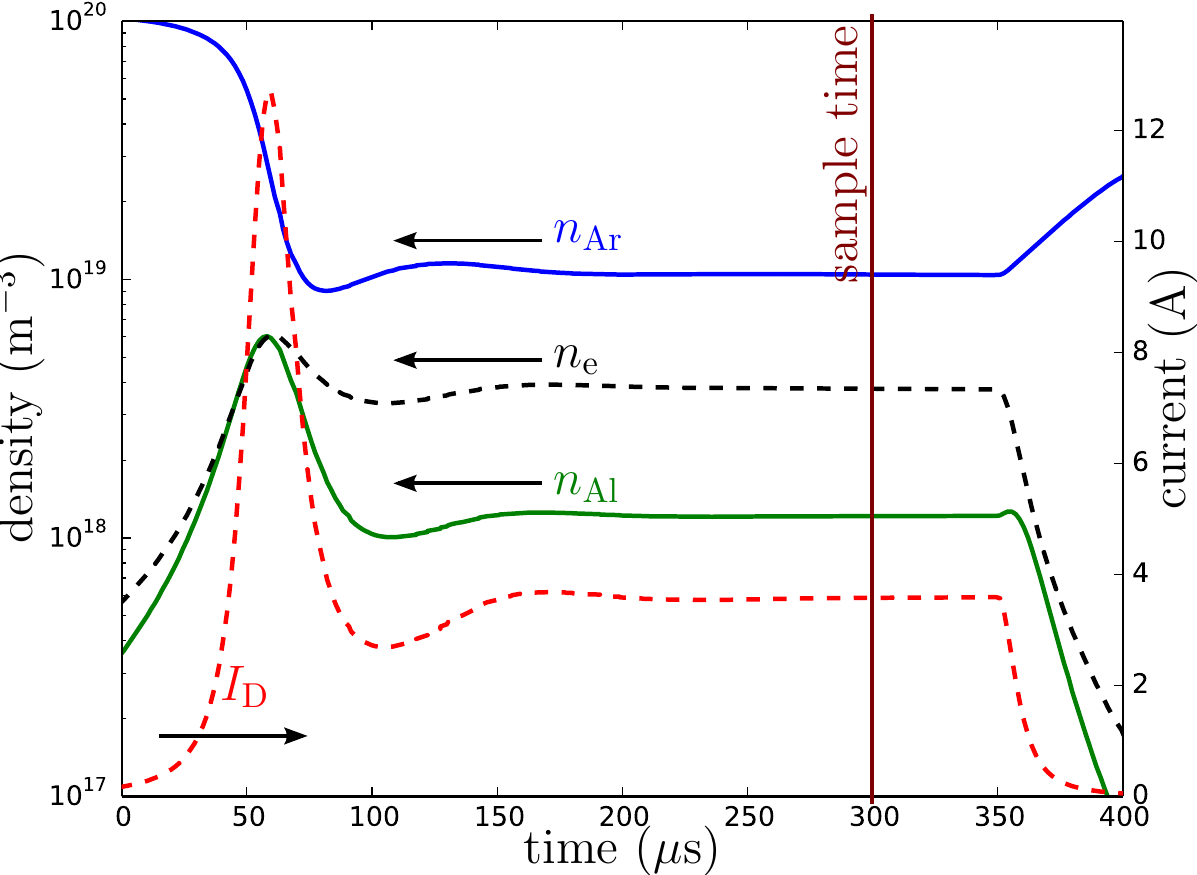}
\caption{Ar (blue), Al (green) and electron (dashed black) densities' evolution in time. The dashed red line shows the discharge current. As the discharge current reaches a plateau, so do the species densities. The Ar is strongly depleted, the Al is initially created by sputtering and subsequently depleted by ionization. The species densities stabilize to the values reported in table~\ref{table:table_1}, sampled at the instant marked with the dark red line.}
\label{figure_1}
\end{figure}

Figure \ref{figure_1} shows the time evolution of some of the species densities in the control volume together with the discharge current (red dashed line). The discharge current shows a peak as the discharge ignites and then settles to a plateau region about 200 $\mu$s into the pulse. The Ar neutral density $n\s{Ar}$ (blue solid line) shows a fast depletion in the initial phases of the pulse, followed by a roughly constant value later in the pulse. The Al density $n\s{Al}$ (green solid line) increases quickly as the Ar ions begin to produce sputtered material, to be then depleted due to ionization collisions and finally settle to a constant value. The electron density (dashed black line) shows a similar time evolution.

\begin{figure}[htb!]
\includegraphics[width=8cm]{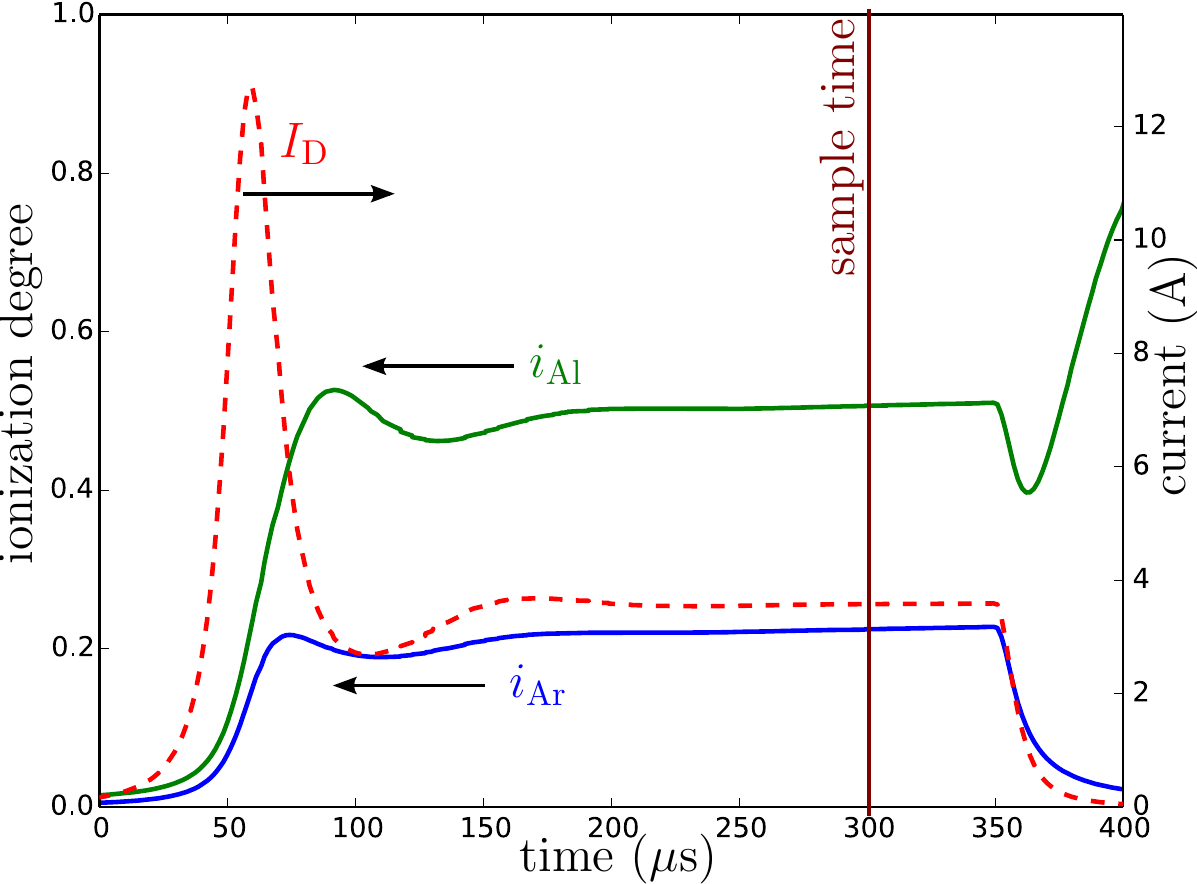} 
\caption{Ar and Al ionization degree (left axis) and discharge current (right axis) against time. The discharge current (dashed red curve) exhibits a plateau region after about 200~$\mu$s. After the same time the degrees of ionization settle to $50\%$ and $20\%$ for Al and Ar respectively. The dark red vertical line represents the instant at which the quantities in table \ref{table:table_1} are sampled.}
\label{figure_2}
\end{figure}

In figure \ref{figure_2} are plotted against time the ionization degrees for Ar $i\s{Ar} = n^+\s{Ar}/(n\s{Ar}+n^+\s{Ar})$ (blue solid line) and Al $i\s{Al} = n^+\s{Al}/(n\s{Al}+n^+\s{Al})$ (green solid line), together with the discharge current (red dashed line). The Al degree of ionization stabilizes at about $50\%$, with the Ar ionization degree at about $20\%$. As soon as the Al is produced by sputtering it is quickly ionized because of its low ionization threshold ($\eps_i \approx 6$ eV), therefore $i_{\rm Al}$ is high even at rather low power densities. 
One can notice that the Ar density shown in figure \ref{figure_1} (solid blue line) drops by an order of magnitude, while $i\s{Ar} \approx 20\%$: in the numerical simulation the Ar is rarefied by the sputtering wind as well as by the ionization process. The Al sputtered from the target, in fact, exchanges momentum with the Ar and effectively pushes the gas out of the ionization region (see e.g. \cite{2011-RaaduAxnaes}). 

\begin{table}
\begin{ruledtabular}
\centering
\begin{tabular}{cc}
	$\text{\qquad}
	\begin{alignedat}[t]{2}
	n\s{Ar} &=&\, 1 \cdot 10^{19} &\,\text{m}^{-3}\\
	n\s{Al} &=&\, 1 \cdot 10^{18} &\,\text{m}^{-3} \\
	n\s{e}  &=&\, 4 \cdot 10^{18} &\,\text{m}^{-3}
	\end{alignedat}$
	&
	$\begin{alignedat}[t]{2}
	V\s{b} &=&\, 360 &\, \text{eV}\\
	j\s{D} &=&\, 0.41 &\, \text{A/cm}^2
	\end{alignedat}
	\text{\qquad}$
\end{tabular}
\end{ruledtabular}
\caption{\label{table:table_1} Physical parameters used in the calculation of the analytical hot electron eedf. The values are extracted from the results obtained by the volume global model at the instant t = 300 $\mu$s, marked with `sample time' in the figures \ref{figure_1} and \ref{figure_2}.}
\end{table}

\begin{figure}[htb!]
\includegraphics[width=8cm]{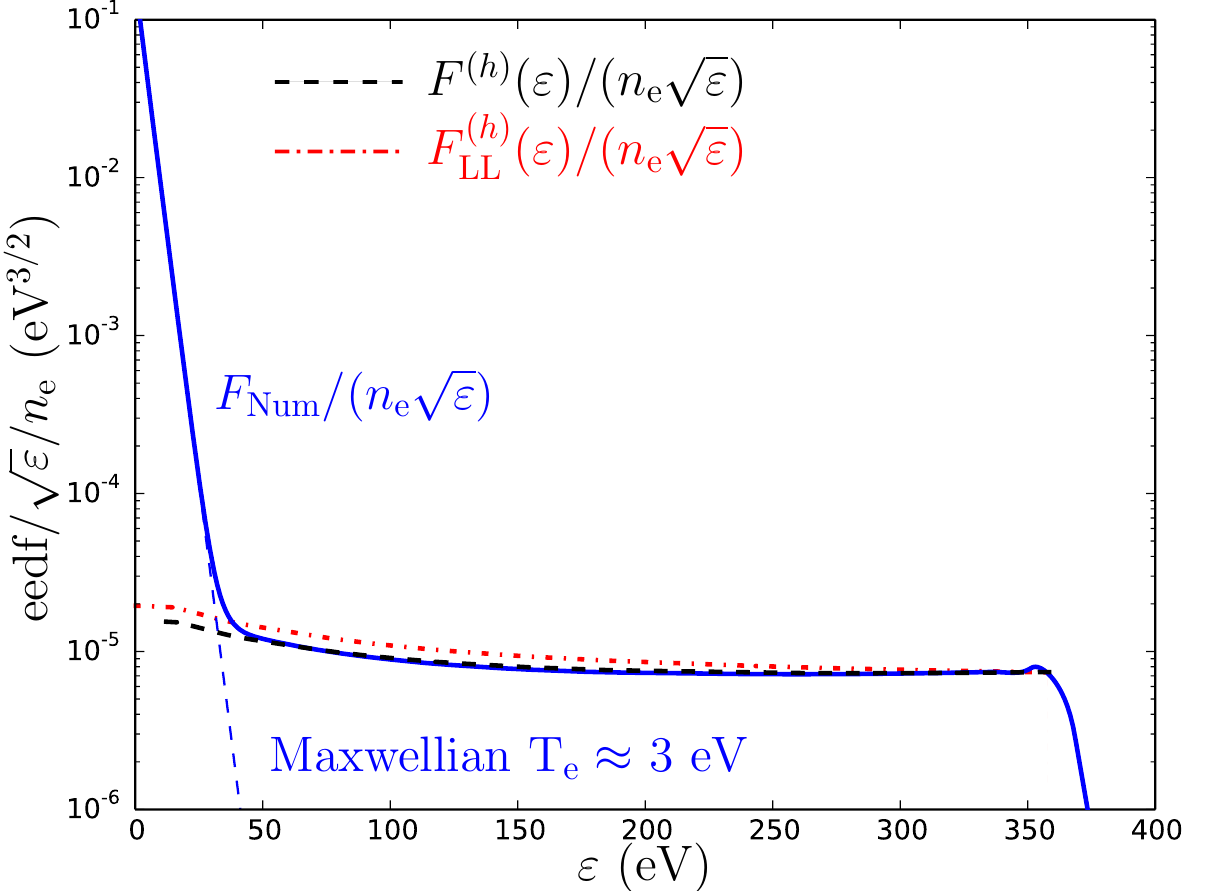}
\caption{Shape of the electron energy distributions against energy from the numerical calculation $F_{\rm Num}$ (blue) at the instant t = 300 $\mu$s, and the analytic solutions, lossless $F\s{LL}^{(h)}$ (dashed red) and $F^{(h)}$ (dashed black). On the vertical axis the distribution functions are normalized as $F/\sqrt{\eps}/n\s{e}$, with [$F$] = m$^{-3}$eV$^{-1}$. Note that the quantity eedf/$\sqrt{\eps}$ is sometimes referred to as the electron energy probability function (eepf).\cite{2005-LiebermanLichtenberg-pg191}}
\label{figure_3}
\end{figure}

\begin{figure}[htb!]
\includegraphics[width=8cm]{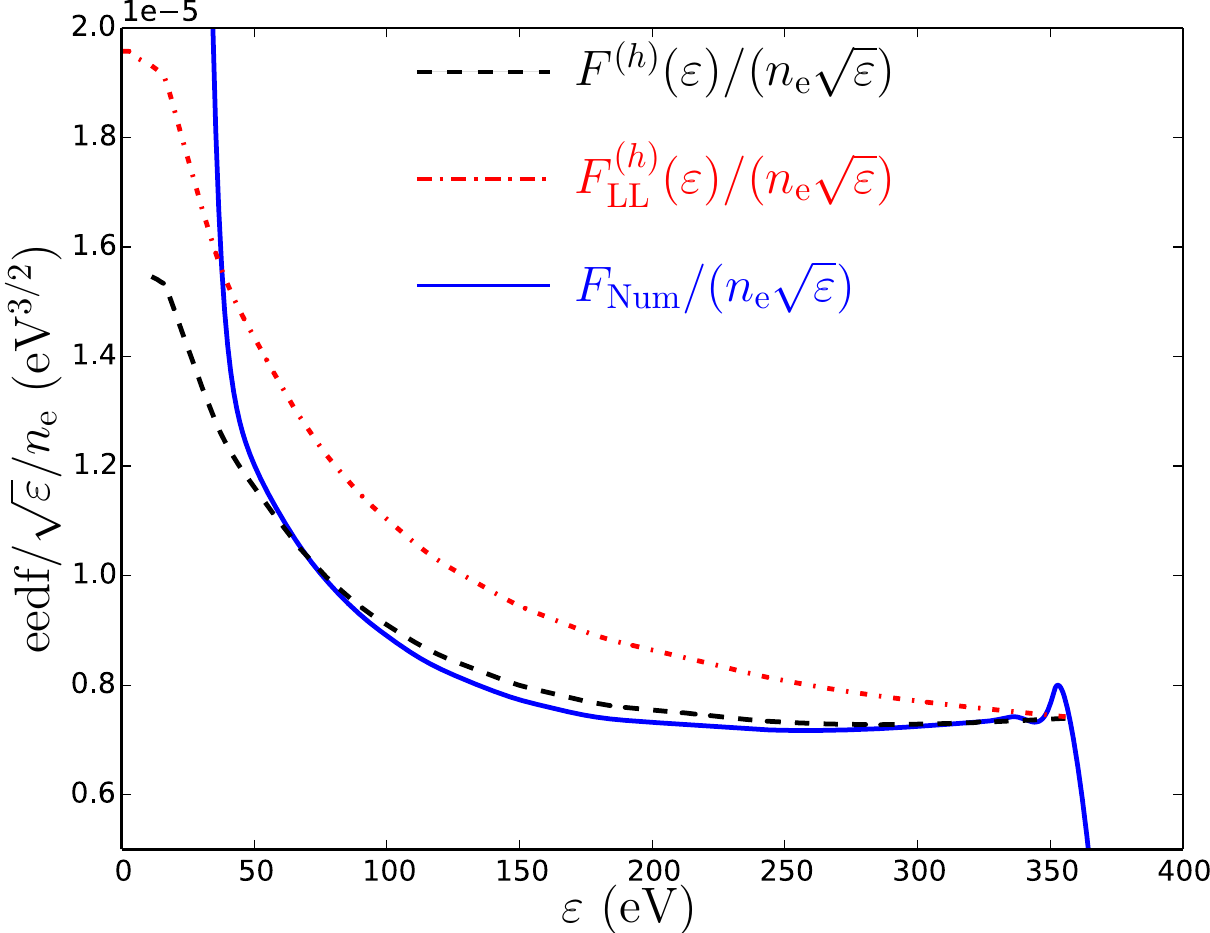}
\caption{Linear zoomed in plot of the normalized electron energy distributions against energy from the numerical calculation $F_{\rm Num}$ (blue) at the instant t = 300 $\mu$s, and the analytic solutions, lossless $F\s{LL}^{(h)}$ (dashed red) and $F^{(h)}$ (dashed black). The agreement of $F^{(h)}$ and $F_{\rm Num}$ in the high energy region is evident. On the vertical axis the distribution functions are normalized as $F/\sqrt{\eps}/n\s{e}$, with [$F$] = m$^{-3}$eV$^{-1}$.}
\label{figure_4}
\end{figure}

The physical quantities during the plateau phase, reported in table~\ref{table:table_1}, are used to calculate the eedf reported in \eqref{Fh} and \eqref{Fh_noLoss}. The shape of $F_{LL}^{(h)}$ (red dashed line) and $F^{(h)}$ (black dashed line) are reported in figure \ref{figure_3}. In the same figure, the blue solid line is the numerical solution obtained by the global simulation. The numerical result shows two well separated populations of electrons: the cold Maxwellian ones, represented with a straight dashed blue line and the energetic electrons. The numerical result shows that the eedf starts to deviate from a Maxwellian at energies of about 30 eV, but that most of the electrons fall into the Maxwellian category: the eedf for the hot electrons is about 4 orders of magnitude smaller. This result supports the claim that the density of the cold or Maxwellian electrons is much larger than the one of the hot population: the hot electron density is less than $5\%$ of the total density. 
As for the analytic solutions, the agreement of $F^{(h)}$ and $F\s{Num}$ is excellent for electrons with energies larger than $\approx 60$ eV, i.e. within the region of validity of the energetic electrons assumption as shown in the zoomed in plot in figure \ref{figure_4}. 
In figure \ref{figure_3} it is interesting to notice the appearance of a plateau in energy which extends from about 75 eV to 350 eV. This feature should occur in general in systems where an initially monoenergetic beam interacts both with neutral species and Maxwellian electrons, provided that energetic electrons have a large energy with respect to the inelastic collision processes threshold and that the confinement time is large compared with the slowing down time. In the case under study, the average confinement time is $\langle 1/\nu^*\s{L}\rangle \approx 20 \ \mu$s, while the slowing down time for the most energetic electrons is $\langle q\s{e} V\s{b}/v_\eps(\eps)  \rangle \approx 5\ \mu$s.

\begin{figure}[htb!]
\includegraphics[width=8cm]{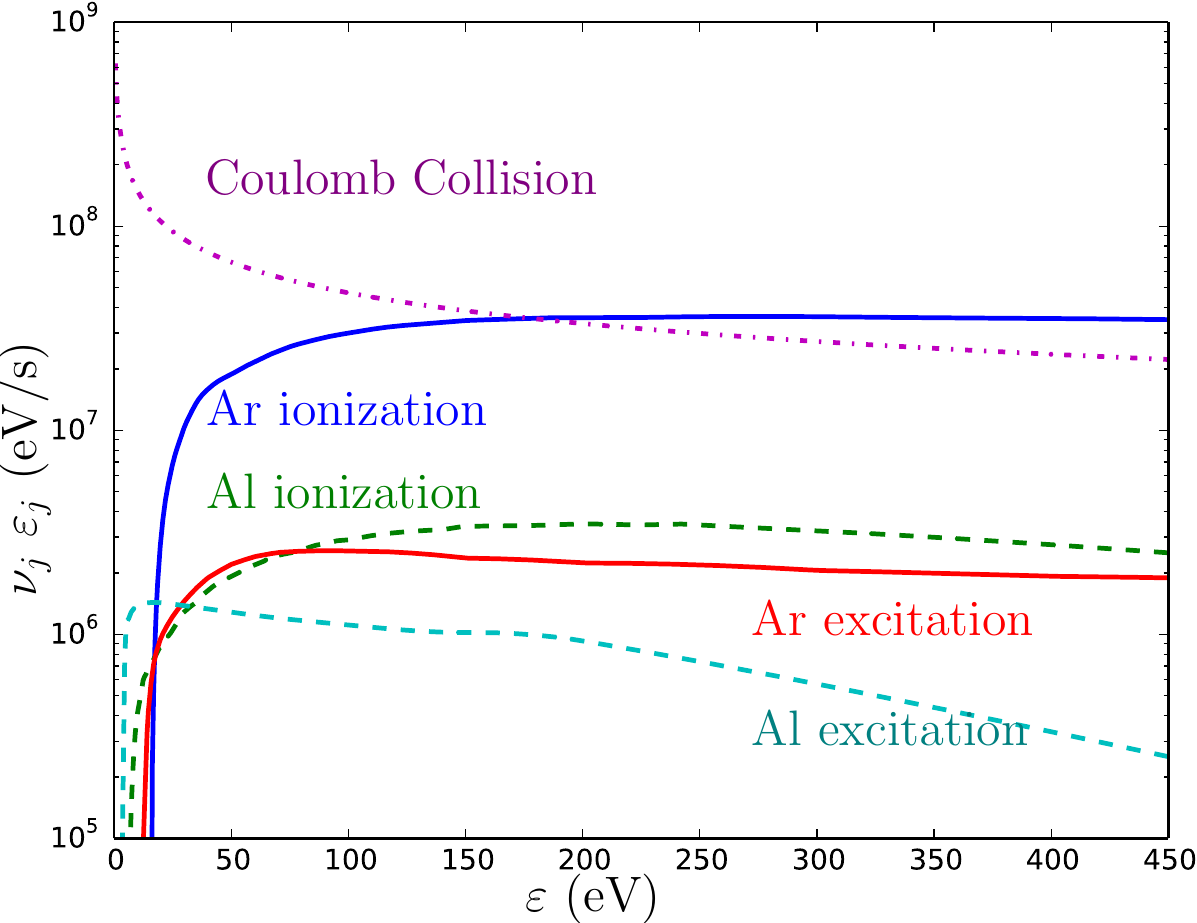}
\caption{Terms of the type $\nu_j \eps_j$ in equation \eqref{veps} against energy of the hot electrons, identified by the collisional process. The densities employed are the ones reported in table~\ref{table:table_1} and obtained at  t = 300 $\mu$s. The solid line identifies Ar, the dashed line Al and the point-dash line Maxwellian electrons. The contributions due to Coulomb and Ar ionization collisions dominate.}
\label{figure_5}
\end{figure}

The lossless solution $F_{LL}^{(h)}$ approximates well, within a $20\%$ error, the more accurate $F^{(h)}$, with an error that increases for lower electron energies as the exponential decay in equation \eqref{Meps} with decreasing energy becomes greater.
Despite the simpler chemistry model employed in the analytic calculation with respect to the numerical global model, the former is able to capture the most relevant processes, and allows for a partial verification of the numerical model. 
Table \ref{table:table_2} reports a list of reactions, with the appropriate reference, and indicates with a checkmark (\checkmark) whether the reaction is included in the analytic or the numerical model. All calculations and the results reported in figures \ref{figure_5}-\ref{figure_7} employ the cross sections from the sources referenced in table \ref{table:table_2}.
Figure \ref{figure_5} shows a plot of the terms of the type $(\nu_j \eps_j)$ using the data reported in table \ref{table:table_1}; these terms determine the drift velocity in energy space \eqref{veps} and therefore the analytic eedf. The Coulomb collisions with the Maxwellian cold background and the Ar ionization terms are here the dominant processes.

\begin{table}
\begin{ruledtabular}
	\begin{tabular}{l l l |c c |c}
\multicolumn{3}{c|}{\multirow{2}{*}{Reaction}} & Analytical  & Numerical\,\,  & \multirow{2}{*}{Ref.\,}\\
& & & model & model & \\
   \noalign{\vskip -1pt}	
	\hline
	\noalign{\vskip 1pt}
	Ar+e 		     & $\longrightarrow$ & Ar$^+$ + 2e   & \checkmark & \checkmark & \,\cite{2003-Hayashi-Ar} \\
	Ar$^*\s{4s}$+e  & $\longrightarrow$ & Ar$^+$ + 2e		  & 		   & \checkmark & \, \cite{1989-Vlcek} \\
	Ar$^*\s{4s'}$+e & $\longrightarrow$ & Ar$^+$ + 2e		  & 		   & \checkmark & \, \cite{1989-Vlcek} \\
	Ar+e 			 & $\longrightarrow$ & Ar$^*$ + e 		  & \checkmark & \checkmark & \, \cite{2003-Hayashi-Ar} \\
	Al+e 			 & $\longrightarrow$ & Al$^+$ + 2e		  & \checkmark & \checkmark & \, \cite{1975-Shimon-AlIon}\\
	Al+e 			 & $\longrightarrow$ & Al$^{++}$ + 2e	  &            & \checkmark & \, \cite{1982-Crandall-Al2p}\\
	Al+e 			 & $\longrightarrow$ & Al$^*$ + e 		  & \checkmark & \checkmark &\, \cite{1982-McGuire}\\
	Ar$^+$+Al 		 & $\longrightarrow$ & Ar + Al$^+$ 		  & 		   & \checkmark & \, \cite{2000-Lu-PennChExc} \\
	Ar$^*\s{4s} $+Al & $\longrightarrow$ & Ar + Al$^+$ 		  & 		   & \checkmark & \, \cite{2000-Lu-PennChExc}  \\
	Ar$^*\s{4s'}$+Al & $\longrightarrow$ & Ar + Al$^+$ 		  & 		   & \checkmark & \, \cite{2000-Lu-PennChExc} \\
	Ar$^*\s{4s} $+e  & $\longrightarrow$ & Ar + e 			  & 		   & \checkmark & \, \cite{1995-Ashida} \\
	Ar$^*\s{4s'}$+e  & $\longrightarrow$ & Ar + e 			  & 		   & \checkmark & \, \cite{1995-Ashida} \\
	\end{tabular}
\end{ruledtabular}
\caption{\label{table:table_2} List of reactions taken into account in the analytic and numerical models.
The last column lists the references from which the cross sections or rate constants were obtained.}
\end{table}

\section{Applicability of analytic solutions}\label{Applicability}
As already mentioned, the analytic result for $F^{(h)}$ has been derived with the aim of describing the hot electrons in an HiPIMS system.  
HiPIMS discharges are generally dynamical systems, but the analytic result for $F^{(h)}$, obtained under the assumption of a stationary regime, can still be employed: the time scale of relaxation of the electron distribution function is of the order of ten microseconds and most of the current pulse characteristics remain constant for longer intervals as shown in \cite{2007-Anders-VICharact}. In particular, in  \cite{2007-Anders-VICharact}, it is shown that for an aluminum target the discharge current remains constant for time intervals larger than a hundred $\mu s$, for a wide range of applied voltages. One can assume that if the discharge current is stationary, the species densities relevant in the calculation of $v_\eps(\eps)$ \eqref{veps} also remain constant.
The mechanism of heating the plasma via energetic electrons is common to both HiPIMS and dcMS discharges. In conventional dcMS systems, the dominant species is the discharge gas, whereas in the HiPIMS case the sputtered metallic species plays an important role as the self sputtering contribution to the discharge is strong. Therefore, provided that the bias voltage is sufficiently high (roughly above 300~V), the analytic solution for $F^{(h)}$ is applicable to both conventional dcMC and HiPIMS discharges. \\
Magnetron systems are a particular type of glow discharges, which are characterized by a low temperature plasma that is sustained by energetic electrons \cite{1977-Thornton-MSreview}. Therefore it is expected that the analysis here presented could be applied to glow discharges, provided that they can be described by a non-local approach. 
For instance, the negative glow of a dc He discharge addressed in \cite{1991-Lawler} could be a candidate for the application of this analysis: the electrons are electrostatically confined in the positive column where the homogeneous assumption can be made. 
In this case, only the Coulomb collisions and the excitation of metastable levels should be considered in equation \eqref{Fhsimpl2}, and the boundary condition discussed in \ref{Bcs} should also be written accordingly, as it determines the introduction of hot electrons at a given energy.
Moreover, the assumption of a monoenergetic beam of electrons entering the volume after the acceleration in the cathode drop to the full sheath potential should be carefully considered, as the sheath thickness is in this case larger.

\section{Study of the beam-bulk interaction terms}\label{SensitivityAnalysis}

\begin{figure}[htb!]
	\includegraphics[width=8cm]{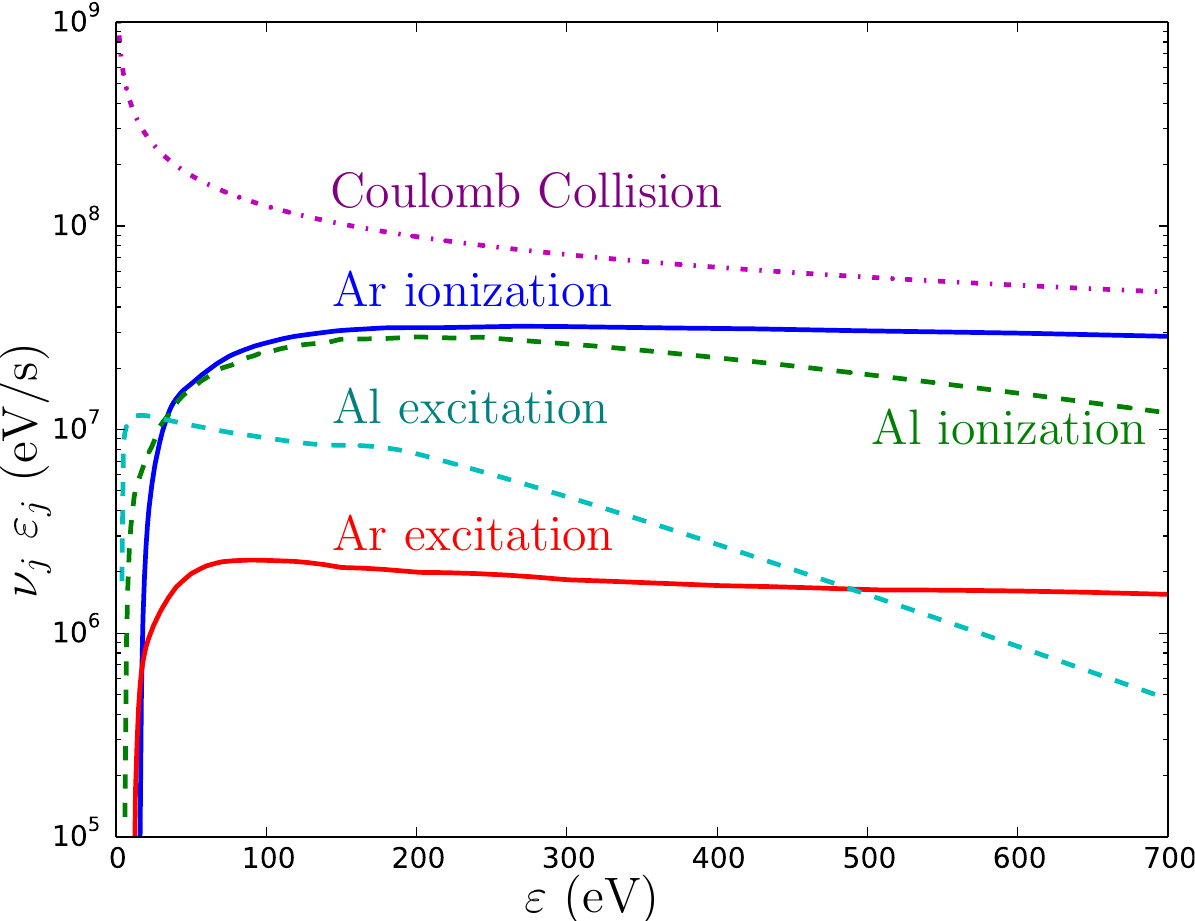}
	\caption{Terms of the type $\nu_j \eps_j$ in \eqref{veps} against energy of the hot electrons, identified by the collisional process. The densities employed are $n\s{Ar} = n\s{Al} = n\s{e} = 10^{19}~\text{m}^{-3}$. The solid line identifies Ar, the dashed line Al and the point-dash line Maxwellian electrons. The contributions due to Coulomb collisions with the Maxwellian electrons dominate.}
	\label{figure_6}
\end{figure}

\begin{figure}[htb!]
	\includegraphics[width=8cm]{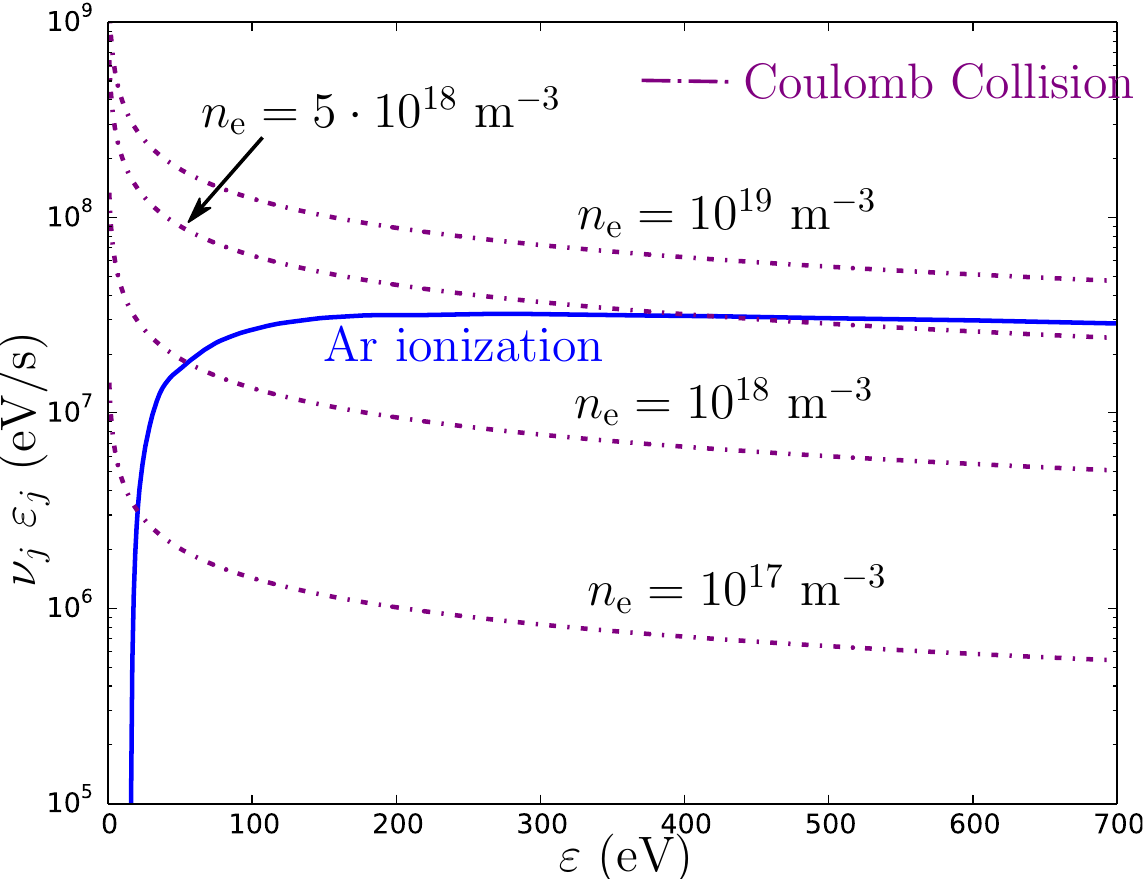}
	\caption{Terms of the type $\nu_j \eps_j$ in \eqref{veps} against energy of the hot electrons, identified by the collisional process. The solid blue line identifies the Ar ionization term with $n\s{Ar}~=~10^{19}$~m$^{-3}$, while the point-dash lines show the Coulomb collision term for Maxwellian electrons. The Maxwellian electron density $n\s{e}$ is varied from $10^{17}$ to $10^{19}$~m$^{-3}$ and the Al density is set to zero.}
	\label{figure_7}
\end{figure}

The velocity $v_\eps(\eps)$ given in equation \eqref{veps} represents the interaction between the hot electrons and the other species, and both analytic solutions \eqref{Fh} and \eqref{Fh_noLoss} depend heavily on its value. On the other hand, $v_\eps(\eps)$ has a large range of variability depending on the neutral species and Maxwellian electron densities. 
Therefore, it is relevant to visualize the relative magnitude of the terms in \eqref{veps} for different values of the densities. 
The purely illustrative figure \ref{figure_6} shows a plot of the terms of the kind ($\nu_j \eps_j$) for $n\s{Ar} = n\s{Al} = n\s{e} = 10^{19}$ m$^{-3}$. When all species have the same density, the Coulomb interaction of the hot electrons with the Maxwellian ones dominates all other processes, followed by Ar and Al ionization.
On the other hand, in most realistic cases, the densities will differ quite significantly: at lower voltages (or during the first phases of the discharge) the Ar species dominates the Al, while in the limit of extreme self sputtering regimes the Al species dominates. As for the electron species, in the high power regime, the ionization degree is expected to approach unity, therefore the Coulomb collision contribution is the most conspicuous. 
Figure \ref{figure_7} shows the sensitivity of $\nu\s{s}^{hM}(\eps)$ to the Maxwellian electron density: $n\s{e}$ is varied from $10^{17}$ to $10^{19}$ m$^{-3}$, 
{while $n\s{Al}$ is set to zero}. 
For comparison, the term $ \nu\s{ion} \eps\s{ion}$ is also plotted as a solid line: it is interesting to notice that the two collisional processes have equal influence when $n\s{Ar} = 10^{19}~\text{m}^{-3}$ and $n\s{e} = 5 \cdot 10^{18}~\text{m}^{-3}$.

\section{Conclusions and outlook}

The analytic solutions $F^{(h)}$ \eqref{Fh} and $F\s{LL}^{(h)}$ \eqref{Fh_noLoss} describe systems that are characterized by a cold Maxwellian bulk and a hot electron population originating with secondary electron emission. 
The cooling of the secondary electrons is carried out by inelastic ionization and excitation collisions with neutral species, and by elastic Coulomb interactions with the Maxwellian electron population. 
This analysis is applicable, among other systems, to dc magnetron sputtering (dcMS) and high power impulse magnetron sputtering (HiPIMS): the secondary electrons emitted by the target reach the ionization region after being accelerated by the sheath potential. In this region close to the target, electrons are magnetically confined and exhibit a spiral like motion around the magnetic field lines while bouncing back and forth in the magnetic mirror. This motion allows the energetic electrons to experience a large number of collisions even at a low pressure (usually below 1 Pa).  
HiPIMS systems differ from dcMS because of the higher peak power density they achieve (several kW/cm$^2$): the bias voltage applied to the target is larger, resulting in a higher discharge current, which is applied in pulses up to a few hundred microseconds. \\
Both analytic solutions have been used to describe a HiPIMS discharge. The species densities used in the expressions \eqref{Fh} and \eqref{Fh_noLoss} are obtained from the numerical solution of a kinetic global model, to be addressed in a follow up publication. The simulated discharge shows a plateau region where both the discharge current and the species densities are constant: the stationary eedf is calculated in this region. 
The analytic result is then compared to the numerical one obtained by the model and shows excellent agreement despite the simplifications made. 
The applicability of the analytic solution together with its sensitivity to the main parameters is addressed in section \ref{Applicability} and \ref{SensitivityAnalysis}. The analytic solution \eqref{Fh} can be approximated by its lossless version \eqref{Fh_noLoss}, provided that the neutral species density is low, the magnetic field is strong and the bias voltage is larger than about 300~eV. For ionization degrees larger than about $50\%$, the Coulomb collision term in the drift velocity in energy space $v_\eps$ dominates the Ar and Al ionization terms.\\ 
The analytic solution of the energetic electron edf can be used to calculate correction terms to be used in a fluid description of the electron species. In fact while the Maxwellian electrons determine the electron density and the ambipolar field, the energetic ones are responsible for the power coupling to the discharge and the majority of the ionization in the negative glow region. Knowledge of the hot electrons' edf allows one to devise a fluid model with corrections for kinetic effects, i.e. a hybrid model. 


\section{Acknowledgment}
This work has been supported by the German Research Foundation (DFG) within the frame of the Collaborative Research Centre TRR 87 'Pulsed high power plasmas for the synthesis of nanostructured functional layers' (SFB-TR 87). The authors gratefully acknowledge fruitful discussions with Dr G\"{u}\c{c}l\"{u}, Prof A Smolyakov, Dr Teresa de Los Arcos and Dr D Eremin.\\

\bibliography{main_Fh}

\end{document}